\documentclass[twoside,11pt]{article}

\usepackage{blindtext}
\usepackage{booktabs}
\usepackage[preprint]{jmlr2e}

\usepackage{lastpage}

\ShortHeadings{}{Predicting Long-Term Student Outcomes from Short-Term EdTech Log Data}
\firstpageno{1}

\begin{document}

\title{Predicting Long-Term Student Outcomes from Short-Term EdTech Log Data}

\author{\name Ge Gao \email gegao@stanford.edu \\
       Stanford University
       \AND
       \name Amelia Leon \email amelialt@cs.stanford.edu \\
       Stanford University
       \AND
       \name Andrea Jetten \email A.m.jetten@fsw.leidenuniv.nl \\
       War Child Alliance
       \AND
       \name Jasmine Turner \email jasmine.turner@warchild.net \\
       War Child Alliance
       \AND
       \name Husni Almoubayyed \email halmoubayyed@carnegielearning.com \\
       Carnegie Learning
       \AND
       \name Stephen Fancsali \email sfancsali@carnegielearning.com \\
       Carnegie Learning
       \AND
       \name Emma Brunskill \email ebrun@cs.stanford.edu \\
       Stanford University
       }

\editor{My editor}

\maketitle

\begin{abstract}
Educational stakeholders are often particularly interested in sparse, delayed student outcomes, like end-of-year statewide exams. The rare occurrence of such assessments makes it harder to identify students likely to fail such assessments, as well as making it slow for researchers and educators to be able to assess the effectiveness of particular educational tools. Prior work has primarily focused on using logs from students full usage (e.g. year-long) of an educational product to predict outcomes, or considered predictive accuracy using a few minutes to predict outcomes after a short (e.g. 1 hour) session. In contrast, we investigate machine learning predictors using students’ logs during their first few hours of usage can provide useful predictive insight into those students’ end-of-school year external assessment. We do this on three diverse datasets: from students in Uganda using a literacy game product, and from students in the US using two mathematics intelligent tutoring systems. We consider various measures of the accuracy of the resulting predictors, including its ability to identify students at different parts along the assessment performance distribution. Our findings suggest that short-term log usage data, from 2-5 hours, can be used to provide valuable signal about students' long term external performance. 
\end{abstract}


\section{Introduction}


Though educational software, including intelligent tutors and educational games, is increasingly ubiquitous, evaluating the effectiveness of such software remains a challenge, at both an individual level and a population level. Typically school districts are most interested in long-term student outcomes, like end of year state or national assessments. Though such assessments are generally highly rigorous and carefully designed, their rare occurrence introduces multiple challenges, making it harder to identify students likely to fail such assessments, as well as making it slow for researchers, educators, product designers and policy makers to be able to assess the effectiveness of particular educational tools. It has long been recognized (\textit{e.g.}, in the broad investigation with ASSISTments~\citep{pardos2006using,pardos2014affective,walonoski2006detection,leon2024estimating}) that the log data generated when students interact with educational software could themselves be used as a form of temporally integrated insight into a student's state of knowledge. In this work we investigate machine learning predictors using students' logs during their first few hours of use with educational technology can provide useful predictive insight into those students' end-of-school year external assessment. 

There are multiple reasons such predictors would be helpful. Those predictors might be used within the software itself, to help introduce different forms of pedagogical instruction and support to students who are struggling, or even new challenges to those that are thriving. While many educational tools already involve some aspects of personalization, to our knowledge, such tools typically rely on proxy measures of student outcomes (such as performance on a set of internally defined skills). Though there is some prior work relating such proxies as observed over the school year to external assessments~\citep{ritter2013predicting,zheng2019using}, it is still unclear if such features as observed over a full year are necessarily predictive given only \textit{a limited window}, which is important if these shorter term signals are used for instructional decisions. 

Equally importantly, that short-term information could be provided to teachers to help them better understand the progress of individual students and their class as a whole, potentially informing the need for additional resources or changes in strategy. For example, a teacher might assign an aide to spend more time with a struggling student, or they might choose to increase the amount of time spent on math if their whole class is likely to perform poorly. 
In general, it is unclear if many educational products and software are carefully aligning short-term observations to make automated instructional decisions, towards maximizing desired long-term outcomes. Indeed, doing so has often been very challenging due to the limited time horizons involved, and to our knowledge there has been limited prior work on using such short horizon log data to predict delayed long term outcomes.

Rather, prior work has tackled other related aspects to the problem we consider. First, there have been a number of papers showing that using long-horizon student data logs can be used to help predict external assessments, across multiple intelligent tutoring systems and educational software products~\citep{joshi2014generalizing,anozie2006predicting,ayers2008irt,ritter2013predicting,feng2006predicting,pardos2006using,zheng2019using}.  For instance, Ritter at al.~\citep{ritter2013predicting} and Zheng et al.~\citep{zheng2019using} used students’ log data, demographics, and pre-assessment scores from an academic year to predict standardized tests/outcome using machine learning methods, and Feng et al.~\citep{feng2006predicting} investigated ASSISTments data through a year to predict a high-stake state test (MCAS) at the end of the year using log data and pre-assessment score. Such work has typically shown that log data provides a useful signal to help predict student tests scores, focusing on population measures like Pearson correlation and root mean squared error.
Recently, perhaps in part motivated by the significant challenges during the covid-19 pandemic of conducting standard educational assessments when many students are remote, people have been interested in designing much shorter assessments that have similar benefits to existing, much longer assessments -- for example, Tran et al.~\citep{tran2023development} developed a reading assessment that calculates each student's score at 10 seconds time intervals which was then correlated against a full 3-minute standardized test scores. In such settings, the assumption is that the assessment is being done to capture static student performance, rather than extracting signal during standard usage of a product designed to support student learning.\footnote{An interesting possibility if very short test become feasible is whether they could be more widely incorporated in standard software, so as not to distract from learning, which is often a challenge with lengthy assessments.}

Most related to our current work is a couple recent papers that have similarly examined whether very short-term data from student log outcomes can predict delayed student outcomes~\citep{ahadi2015exploring,liao2019robust,castro2017evaluating,gao2021early,emerson2019predicting,mao2019one}. However, in those settings, the length of horizon was very short, focused on a single session of interaction with a student. For example, Gao et al.~\citep{gao2021early} examined how performance on the first problem related to post-test after 5 problems, and Mao et al.~\citep{mao2019one} looked at performance on the first 1 minute related to the final outcome after 20 minutes. 
While such work has a related motivation, in contrast, in our work we are interested in much longer time scales, seeking to use a limited amount of likely multi-session data to predict external assessments taken months later to evaluate student learning (as well as the impact of educational support) on a much broader scale. In addition, to our knowledge, prior research that has developed predictors of external student assessments using many session (e.g. school year) student tutoring logs, has analyzed performance on a single educational tool and/or platform, leaving open trends and similarities across educational technology systems. 

We note that there has been significant work on developing surrogate measures of delayed outcomes in social science and economics~\citep{athey2019surrogate,zhang2023evaluating,hohnhold2015focusing,gao2022reinforcement}. Such approaches generally built a model to predict long-term outcomes from a set of short-term outcomes, and estimated long-term treatment effects using the predicted long-term outcomes. And prior works have found that leveraging the short-term observed information from humans can provide reliable estimation for long-term outcomes. For example, Athey et al. predicted employment many years after a short-term job training program, using a surrogate of 1 year employment status~\cite{athey2019surrogate}. Zhang et al. used 14 days of users' data to develop a surrogate index, which was highly correlated to a directly measure of 63-day treatment effects~\citep{zhang2023evaluating}. Surrogate endpoints are also used in clinical settings, when the desired outcome of interest may be substantially delayed (such as 5 year survival rates) and other shorter term measures are known to be predictive of the long-term outcome.\footnote{https://www.fda.gov/about-fda/innovation-fda/fda-facts-biomarkers-and-surrogate-endpoints}, and increasingly in other settings 
such as finance and recommendation systems~\citep{wang2022surrogate,zhang2023evaluating,hohnhold2015focusing}. Those are potentially related to our interests, and motivate us to developing models by using short-term log data for estimating long-term external outcomes given an educational system.       
 
When predicting students' long-term performance, another challenging situation is that available observations of students can be commonly limited. Prior works have found some powerful indicators (e.g., demographics, pre-test scores, knowledge components (KCs), etc.)~\citep{acharya2014early,ritter2013predicting,zheng2019using,alhazmi2023early} and expert-defined features (e.g., clicker questions, programming error/distance metrics)~\citep{liao2019robust,porter2014predicting,jadud2006methods,carter2017using,watson2013predicting,piech2012modeling} to predict students' outcomes at an early stage. However, those features may not always be available to tutors. And most of the prior works mainly focused on a specific platform or context, there is lack of investigation on what features might be generally important across contexts to guide tutors from a new context to quickly understand the potential future outcomes of students. Moreover, while a crucial goal for developing new tools and interventions is broadly enhancing learning outcomes over student populations, it is also pivotal to understand its potential effects on varied students subgroups sorted by performance at an early stage, since we would like to understand whether various performers could benefit from new interventions. However, prior work mainly focus on developing techniques to enhance prediction on specific subgroups (\textit{e.g.},~\citep{helal2019identifying}), while we are further interested in evaluating over both population and subgroups during predicting long-term outcomes, with respect to short horizon and the sets of predictive features, using features that may generalizable across contexts.      


In this work, we investigate the prediction of long-term, external students' outcomes using their short-horizon log data, across tutoring systems across three different educational contexts: the Can't Wait to Learn reading educational technology games (using data from students in Uganda), iReady middle school math intelligent tutor (used by seventh graders in the United States) and MATHia middle school math intelligent tutor (used by 6-8 graders in the United States). We explore the potential of using short horizon features that could be generally extracted from log data across contexts, without extensive domain knowledge of the underlying educational software tool or student demographics or student prior performance data. We do this in part as such data may not always be available for many reasons, including the common case of new students who transfer to a new district, particularly midway through an academic year. In addition, by focusing on broadly similar features that are likely to be present across many educational platforms, we can evaluate the similarities and differences across settings. Using such features, we compare the performance of three popular machine learning models (\textit{i.e.}, linear regression, support vector regression, and random forest) with respect to the length of horizon and contexts. 

While prior work has primarily focused on population level metrics, part of the motivation for such work is the potential to help support students who are expected to have major challenges months later on an external assessment if their trajectory continues, or to potentially celebrate or challenge a student that is already showing signs of strong expected future performance. To investigate this further, we also analyze the resulting quality of the short-horizon estimates over subgroups of students that are sorted by performance for a more thorough understanding about the prediction performance. Moreover, we examine the effects of pre-assessments on predictions over both population and subgroups, with different set of features, to understand if pre-assessments or pre-test data, when available, is similar, different or complimentary, to student log data. 
Specifically, we explore the following research questions:
\begin{enumerate}
    \item Can we consistently, across multiple educational tools, obtain a short-horizon log data only predictor of student long-term outcomes which provides significant predictive power on external, much delayed educational assessment? How does its accuracy compare to using the full log horizon data, and does performance versus horizon  vary by datasets/settings?
    \item Does the machine learning algorithms used to form the predictors significantly impact the resulting accuracy of the external assessment predictor?
    \item Is there a stable set of log data features across domains and datasets that is needed to form accurate short-horizon predictors of long-term outcomes? Are multiple features important for improving accuracy? 
    \item What is the resulting quality of the short-horizon estimates and how does it differ across subgroups of students sorted by performance? 
    Are we systematically better or worse at predicting higher/lower performers?
    \item When pre-assessments are available, is the quality of using log data equivalent to this form of information, and do we see additive gains of combining both? 
\end{enumerate}

\section{Design Context \& Methods}

\subsection{Data}

\texttt{Can't Wait to Learn} (CWTL) is an educational program developed by the nonprofit organization War Child to support children impacted by or residing in conflict affected areas who may have challenges accessing quality education. The educational technology is a curriculum aligned, self-paced, autonomous learning program that aims to teach foundational numeracy and literacy skills, and prior and ongoing work suggests that it can have a positive impact on student learning outcomes (e.g. ~\cite{brown2023can}). The program is delivered on a tablet and targets learning objectives from grades 1-3. Based on the context, the program can be used as a standalone or a supplemental educational program. CWTL has been used in many settings, including South Sudan, Sudan, Lebanon, Jordan, Chad, Bangladesh, Ukraine and Uganda. 

In this paper we used CWTL data collected across 30 schools from September 2021 to December 2022. We have access to log data from 739 students learning literacy (referred as \texttt{CWTLReading} in this paper), who participated in the study with both pre- and post-test scores. The post-test scores were calculated by averaging of sub-task  scores, including letter knowledge, phonemic awareness, reading fluency, and reading comprehension in an assessment.

\texttt{MATHia} (previously Cognitive Tutor~\cite{ritter2007cognitive}) is an intelligent tutoring system for middle school mathematics. The anonymized \texttt{MATHia} data used in this study was collected from 2644 students using the software from August 2020 to June 2021 in a mid-western US state.
The predicted post-test score is assessed by a state end-of-year test. The test has both an integer score, and 5 achievement categories: levels 3 to 5 correspond to a passing score.


\texttt{iReady} is an online program for K-8 reading and/or mathematics that provides an adaptive diagnostic, and online instruction. The \texttt{iReady} data used in this study contains 428 Grade 7 students in mathematics courses with complete records including both pre- and post-test scores from August 2022 to June 2023. The predicted post-test score is assessed by the Smarter Balanced Assessment System (SBAC) which is a standardized test consortium and creates Common Core State Standards-aligned tests to be used in multiple states of the US.  The students were part of another research study with a particular focus on schools with more struggling students.

\subsection{Data Analysis}
In this section, we will outline our approach for estimating delayed external assessments using short-term log data as students work on an educational technology product. We will overview the features extracted from log data, before describing the machine learning algorithms used, and the evaluations of our proposed methods

We first note an important detail, which is how to define usage time. Though it may seem trivial, defining usage time often requires several decisions, including whether to define things as available usage time (the time in a classroom a student had the opportunity to use educational software), the wall clock time a student was logged in to an educational software tool, as well as the active time a student spent doing problems. The first definition (potential usage time) is often very hard to know, since it requires information around attendance. The second one (wall clock or logged in time) is easier to extract for most systems that include a session variable, though not all software systems do -- in particular, our CWTL data did not include a session variable. Even in other systems this can be a challenge as it is common for some software products to automatically log out a student given a sufficiently long inactive time period. For example, if a student logs in at 9:10am, and then is logged out at 9:25am, and then logs in again at 9:40am, and logs out for the day at 10:00am, one could count this as a 50 minutes of usage, or 35 minutes. The 50-minute view might reveal important aspects of disengagement that would be missed if it was treated as 35 minutes. For the \texttt{MATHia} and \texttt{iReady} data, for which we do have session identifiers, we consider wall clock time spent within each session. In contrast, for CWTL, which lacks session identifiers in our available data, we instead accumulate active time on each played minigame. This is possible because the start and end time on each minigame is recorded. Note that in general this may underestimate the total amount of time spent by a student, since it does not include watching videos or time between completing activities. In our discussion we will further reflect on the potential impact of our particular choices around usage time definitions.

\subsubsection{Features Extraction}
While it is possible to make direct predictions from raw click-stream log data, we choose, for interpretability and comparison across multiple educational platforms, to take the common approach of first employing a pare-processing feature extraction step. Briefly, we focused on log features that are generally obtainable but skills or knowledge components may sometimes be not possible. We extract features that are broadly used in prior research and may commonly shared within log data across learning systems (see \textit{e.g.}~\cite{ritter2013predicting,prihar2023effective,zhang2019early}).
In particular, the log features were:
\begin{enumerate}
    \item Aggregate features across the full period of time considered: 
    
    \textit{num\_problem}: the total number of attempted problems; 
    
    \textit{num\_success\_problem}: the number of success problems
    
    \textit{perc\_success\_problem}: the percentage of success problems
    
    \item The three educational technology products considered all have subdivisions into lessons/ workspaces / sections, as do many educational software systems. Here we consider features normalized per section: 
    
    \textit{min\_attempts\_per\_problem}: the minimum number of attempts within a problem
    
    \textit{avg\_attempts\_per\_problem}: the average number of attempts within a problem
    
    \textit{max\_attempts\_per\_problem}: the maximum number of attempts within a problems
    
    \textit{num\_guess\_in\_problem}: the number of likely guess attempts (\textit{i.e.}, attempts accomplished within 2 seconds~\citep{beck2013wheel}) across problems
    
    \textit{num\_idle\_in\_problem}: the number of attempts spent where the time spent for that event was greater than average time across students
    
    \textit{num\_twice\_avg\_time\_in\_problem}: the number of attempts   where the time spent for that event was greater than twice of average time across students

    \textit{num\_long\_idle\_in\_problem}: the number of attempts spent where the time spent for that event was greater than  five minutes 
    
    \textit{avg\_time\_per\_problem}: average time (in seconds) spent on each problem
    
    \textit{avg\_time\_per\_success\_problem}: average time (in seconds) spent on each success problem
    
    \textit{avg\_time\_per\_failed\_problem}: average time (in seconds) spent on each failed problem
    
    \item  Behavior-related features from log data that relate to indicators of wheel spinning and unproductive persistence~\citep{beck2013wheel}:
    
    \textit{num\_unproductive\_persistence\_thres\_5}: the number of consecutive five attempts without succeeding a problem
    
    \textit{num\_unproductive\_persistence\_thres\_10}: the number of consecutive 10 attempts without succeeding a problem
    
    \textit{time\_first\_unproductive\_persistence}: the time step that the first \textit{num\_unproductive\_persistence\_thres\_5} occurs
    
\end{enumerate}

There is considerable variability across educational systems.
For 'problem' level features, our focus is investigating features with respect to the most fundamental level recording activities of each student. For example, in \texttt{CWTL-Reading}, students play educational mini-games, and play multiple bubble games under each mini-game, where they have to answer a minimum number of instances (\textit{e.g.}, 8 out of 10) for the current bubble games. In this work, we refer the bubble game for \texttt{CWTL-Reading} as a 'problem', given each activity recorded in log data is associated with the attempt on the level of bubble game. In \texttt{MATHia}, students learn by solving multi-step problems. In \texttt{iReady}, students learn with a personalized sequence of lessons within each session, and provide response at some time points during each lesson, where we refer the lesson as a `problem' in this work.   

The above listed features can all be considered basic counting features based on prior suggestions from experts. An importance affordance of educational data mining methods is to be able to go beyond what experts have identified~\citep{gao2021early,mao2020time,grover2017framework,wang2022surrogate}. To do so, we also automatically extracted potentially useful behavioral features by using a held-out dataset, which is the training set in the cross-validation. More precisely, we automatically extract sequences of students' activities as temporal features using a sequential pattern by Gao et al.~\citep{gao2021early}. We first extract initial frequent sequential patterns within each outcome group (where we divide students into two outcome groups by split in the median of their post-test scores). Then we use the chi-2 test to identify patterns which are significant frequent across students within one outcome group, and select the top 10 patterns as the features to be used in the prediction, where each feature represents whether its associated pattern occurs within each student. Therefore, the overall number of features used in this study would be 26 (or a value between 16 and 26 if less than 10 patterns are identified).




In addition to the features described above, we also consider using pre-test scores when available. It is of interest to build predictions both with and without pre-test scores, as these may not always be available.

\subsubsection{Predicting Outcomes using Machine Learning Methods}

We then use the above features as input to machine learning algorithms to predict students' external assessments. We consider three standard but popular machine learning models: linear regression (LR), support vector regression (SVR) and random forest (RF)\footnote{We also explored using Gradient Boosted Decision Trees and found GBDR achieved similar results. In particular, we trained GBDR for the Mathia dataset to predict post test performance, and the resulting was RMSE 0.119 when using short horizon (2hr) log data, and 0.116 RMSE for full horizon log data}. We also include a baseline, \textit{i.e.}, using mean scores of final outcomes from training set as the predicted values for test set, for comparison with the prediction models. To make the results more comparable across the three educational tools and corresponding datasets, we project post-test scores to be within\footnote{We also explored z-centering the data but found it made little difference.}[0,1]. We reported the best results from grid searching among C=\{0.1, 1, 10, 50, 100\}, epsilon=\{0.01, 0.1, 1\} for SVR (with kernel=`rbf'), and maximum depth=\{2, 5, 10, 12, 15\} for random forest. Other hyper-parameters followed default settings by Python \texttt{sklearn} package.







\subsubsection{Model Evaluation}
We train the models and report evaluation results using 5-fold cross validation. To measure the performance of prediction models, we report rooted mean squared error (RMSE); (2) the square of the Pearson correlation coefficient R (we report the coefficient of determination, $R^2$). 

\subsubsection{Prediction Quality by Student performance}

Prior work has mostly focused  on reporting average accuracy or correlation between predictions and true outcomes over the entire population (e.g.~\cite{ritter2013predicting,prihar2023effective}). However, one of the key motivations for short horizon prediction of long term external outcomes is the potential to be able to identify when a student or set of students is in need of additional support, or potentially to challenge and congratulate those that are thriving. To do so, we consider how the predictive accuracy varies on student subgroups with different post test performance in two ways. First, we subdivide students into five quantiles  (\textit{i.e.}, Q1, Q2, Q3, Q4, Q5, in ascending order of post-test scores) and assess the models' accuracy with respect to each sub-group. Second, a particularly salient classification may be whether a student is considered to be proficient or above, or not, on the final assessment. We also consider this in the context of the Mathia dataset, and report results below.





\section{Results}

\subsection{RQ1: Short-Horizon Log Data for Predicting Long-Term Outcomes across Multiple Domains}

\begin{figure*}
    \centering
    \includegraphics[width=\linewidth]{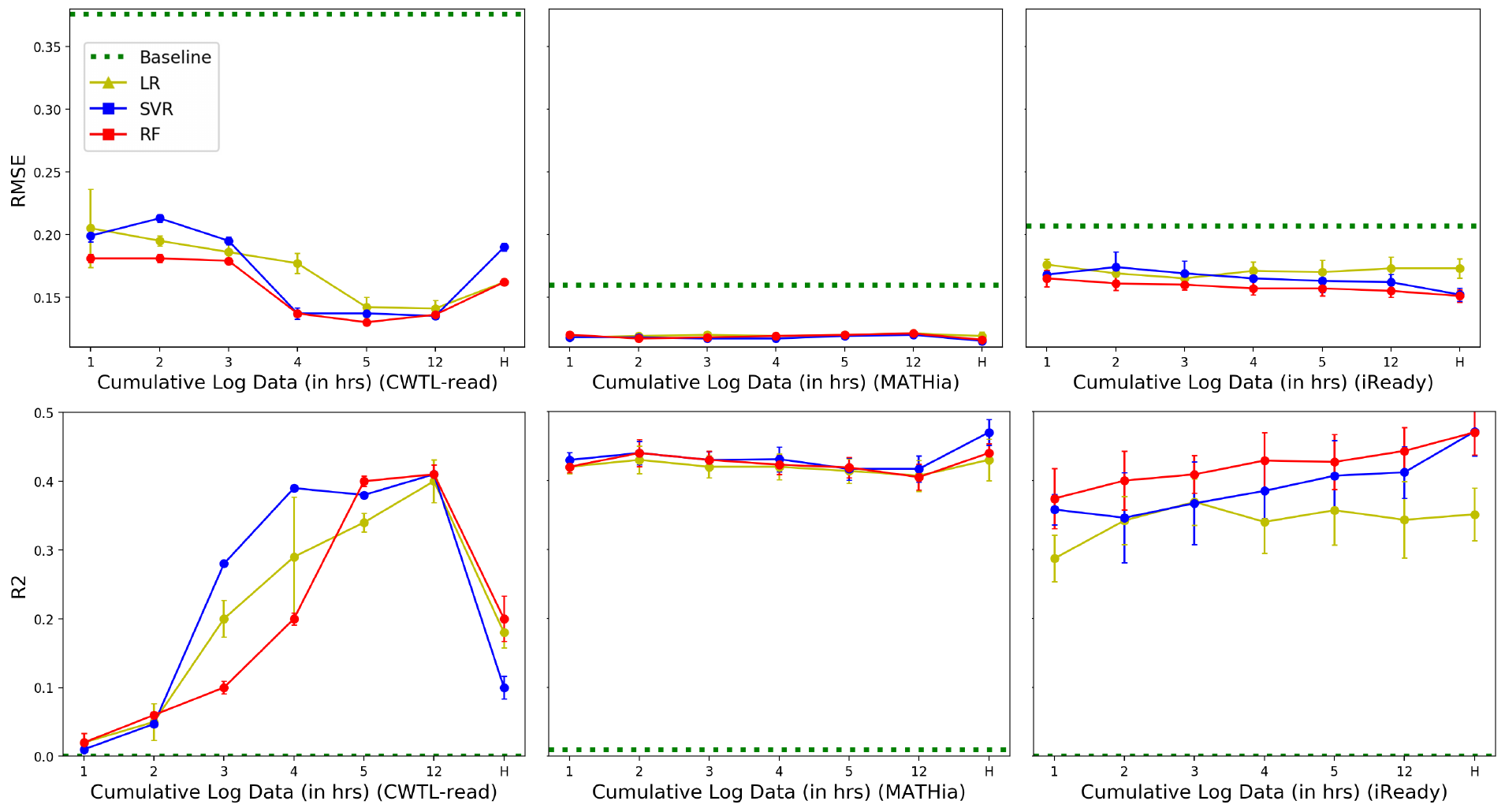}
    \caption{Prediction performance (mean$\pm$standard error from 5-fold cross-validation) with cumulative log data in varied hours across datasets ($H$ is the horizon of the dataset, indicating entire log data is used for prediction). LR: linear regression; SVR: support vector regression; RF: random forest; Baseline: using mean post-test scores from training set as predicted scores. Note all figures on a row share the same y-axis scale, the x-axis is not evenly spaced (see tick marks and labels), and these models only use log data, no additional student demographic or preassessment data.}
    \label{fig:perf_varied_hr_all}
\end{figure*}

To understand the potential signal from using student log data given various lengths of usage since the start of data collection, we train and test each machine learning prediction model with cumulative log data on varied lengths of horizons, \textit{i.e.}, 1, 2, 3, 4, 5, 12 hours, as well as the full length of data (denoted as $H$). The prediction models' RMSE and $R^2$ are presented in Figure~\ref{fig:perf_varied_hr_all}. The x-axis represents the length of horizon for used cumulative log data to train models and y-axis represents the prediction results averaged from 5-fold cross validation. 

Interestingly, we observe that for all three educational products and datasets, there exists a machine learning model built using short-horizon of log data that is nearly as or equally effective as a the best machine learning model built using the full horizon log data. 

The second thing to note is that the short horizon time period with the best performance (as measured by RMSE and $R^2$ varies slightly by datasets/setting. For example, for \texttt{CWTLReading}, the random forest machine learning model has notably better performance with 5 hours of interactive usage (0.13 RMSE and 0.34 $R^2$) but worse performance using log data from 12 hours and full log data. 
In \texttt{MATHia}, two hours of log data is the best of the early horizon (up to 5 hours) exceeds the performance of the full horizon log data, and for \texttt{iReady} strong early performance (measured by $R^2$ and RMSE) among the first 5 hours is obtained at 3 hours, though the performance is quite similar over teh full range. 


We also note that performance not always monotonically improve with longer horizon log data-- machine learning predictors for both \texttt{CWTLReading} and \texttt{MATHia} seem to slightly decrease in performance at the longest horizon lengths.
It is known from the economics  surrogate literature that using a surrogate measure, when surrogacy is satisfied, can result in a lower variance estimate~\cite{athey2019surrogate}, since additional data beyond the surrogate window can introduce additional noise. Though we would not expect to surrogacy to necessarily always hold in real educational data, this may be a reason for why we see, sometimes, higher accuracy estimates using a shorter horizon. 

\subsection{RQ2: Impacts of Selected Machine Learning Algorithms on Resulting Accuracy}
We next assess if the machine learning algorithm used has a significant impact on resulting performance, which we visualize in  
Figure~\ref{fig:perf_varied_hr_all}. In general, all machine learning algorithms are quite similar. For \texttt{CWTLReading}, random forest (RF) achieves the lowest RMSE across different horizons, and achieves the highest overall  $R^2$. Support vector regression (SVR) were slightly less consistent, 
and linear regression (LR) performs relatively more stable across lengths of horizon and performs better than random forest at some early horizons in terms of $R^2$. For \texttt{MATHia}, RMSE for all three models (LR, SVR, RF) is similar, and the difference of $R^2$ between the three methods is minimal on given short horizon log data. For \texttt{iReady}, random forest consistently achieves the lowest RMSE and highest $R^2$, followed by support vector regression, and then linear regression.

All differences are relatively minor, though random forest overall seems to have a slight noticeable benefit in the \texttt{CWTLReading} and \texttt{iReady} datasets. The baseline of simply predicting the average score in the training data performs poorly  across all horizons across datasets.


We also note the predictive models have lower accuracy in  \texttt{CWTLReading} than the other two datasets. A potential reason is that the records across students of \texttt{CWTLReading} are relatively more sparse than the other two datasets. On average, the numbers of logged events over the best short and full data are 270 (5hr) and 1245 for \texttt{CWTLReading}, 586 (2hr) and 8738 for \texttt{MATHia}, 1162 (3hr) and 5042 for \texttt{iReady}. 
This likely merely reflects the differences across different educational technology designs and instrumentation, and it may have an impact on the density of data available for prediction.

\subsection{RQ3: Generalizability of Important Log Data Features across Domains for Long-Term Outcomes Prediction}

\begin{table*}[hbtp]
  \caption{Top 5 important features selected by random forest with 100 decision trees trained on each dataset using short horizon and full log data. `X' represents the feature is selected as the top 5 for the data using a length of horizon. `*' represents the feature is selected as the top 1.}
  \label{tab:shared_feature}
  \resizebox{\textwidth}{!}{\begin{tabular}{|l|ll|ll|ll|}
  \toprule
Feature & \multicolumn{2}{l}{\texttt{CWTLReading}} & \multicolumn{2}{l}{\texttt{MATHia}} & \multicolumn{2}{l}{\texttt{iReady}} \\
& Short Horizon (5hr) & Full Horizon & Short Horizon (2hr) & Full Horizon & Short Horizon (3hr) & Full Horizon \\
  \midrule
Perc. Success Problem & X* & & X* & X* & & X  \\ \hline
Avg. Attempts per Problem & X & X* & X & & X* & X* \\ \hline
Num Idle in Problem & X & X & & & & \\ \hline
Num Problem & X & & & & & \\ \hline
Num Long Idle in Problem & X & X & & & & \\ \hline
Max Attempts per Problem & & X & & X & X & \\ \hline
Num Success Problem & & & X & & & \\ \hline
Num Guess in Problem & & & X & & & \\ \hline
Avg. Time per Success Problem & & & X & X & X & X \\ \hline
Avg. Time Failed Problem & & & & X & & \\ \hline
Time First Unproductive Persistence & & X & & X & X & X \\ \hline
Num Twice Avg. Time in Problem & & & & & X & \\ \hline
Min Attempts per Problem & & & & & & X \\  \bottomrule
  \end{tabular}
  }
\end{table*}




Using interpretable features from log data for our machine learning predictors has the potential to capture interactions that reflect underlying learning processes of students, and allow us to understand how such processes may be similar or different across platforms and temporal periods. 
Towards such insights, we first identify features that are important in our machine learning model, and then also consider if we can use such features to formulate a much simpler model which may replace our more complex machine learning models. 

\textbf{Important features shared across educational contexts.} To quantify important features, we consider top 5 features selected by a  random forest with 100 decision trees (with maximum depth as 10) trained on each dataset, both using a selected short horizon and full data. The results are shown Table~\ref{tab:shared_feature}. Overall, there is a significant overlap across contexts, and across both horizon lengths shown within each context. Specifically, across all three datasets,  
the percentage of times the student succeeded at a problem, and the average number of times a student attempted a problem, 
are frequently selected as important features, in some cases being the most important feature for a dataset. We also note that within a dataset, the important features often significantly overlap between the short and long term important features. For example, the average number of attempts per problem and counts of when a student took a long time to complete a problem, are top features in \texttt{CWTLReading} using both short (\textit{i.e.}, 5 hours) and full horizon. Two features, the percentage of times the student succeeded at a problem and the average time to finish a successful problem, are selected in the top 5 features on \texttt{MATHia} using both short (\textit{i.e.}, 2 hours) and full horizon. Three features, including the average number of times a student attempted a problem, the average time to finish a successful problem, and how long a student persisted unsuccessfully to finish a problem, are selected in the top 5 features on \texttt{iReady} using both short (\textit{i.e.}, 3 hours) and full horizon. The features are generally consistent across datasets and horizon settings, suggests that one might build a general predictive model across educational contexts using the same feature set.

\textbf{Prediction effects using single feature vs. the entire set of features.} In part motivated by the prior observation in the last paragraph, we investigated the effectiveness of using a single feature compared to using a set of features extracted from log data. We train linear regression using the percent success per problem, and the average attempts per problem, respectively, using short horizons as noted in prior sections (\textit{i.e.}, 5 hours for \texttt{CWTLReading}, 2 hours for \texttt{MATHia}, 3 hours for \texttt{iReady}). Table~\ref{tab:single-feature} shows the comparison results in terms of RMSE and $R^2$. Our results show that while a single feature machine learning model still performs much better than the baseline predicting the average performance in a held out set, our machine learning models using the full set of available log features outperform it, in terms of both RMSE and $R^2$. This suggests more complex machine learning models can offer a useful advantage. 

\begin{table*}[hbtp]
  \caption{Performance of LR trained using single feature (\textit{i.e.}, \textit{perc\_success\_problem} (PS) and \textit{avg\_attempts\_per\_problem} (AA)) compared to using entire set of log data features (Entire) on the three datasets using short horizon.}
\label{tab:single-feature}
 \resizebox{\textwidth}{!}{{\begin{tabular}{llllllllll}
  \toprule
  & \multicolumn{3}{l}{\texttt{CWTLReading}} & \multicolumn{3}{l}{\texttt{MATHia}} & \multicolumn{3}{l}{\texttt{iReady}} \\
     & PS & AA & Entire & PS & AA & Entire & PS & AA & Entire \\
     \midrule
     RMSE & 0.177 (0.007) & 0.177 (0.007) & 0.142 (0.018) & 0.123 (0.004) & 0.138 (0.006) & 0.12 (0.005) & 0.197 (0.012) & 0.179 (0.01) & 0.165 (0.012) \\ 
     $R^2$ & 0.1 (0.05) & 0.03 (0.01) & 0.34 (0.194) & 0.39 (0.034) & 0.232 (0.023) & 0.43 (0.053) & 0.074 (0.053) & 0.236 (0.079) & 0.369 (0.076) \\
    \bottomrule
  \end{tabular}}
  }
\end{table*}

\subsection{RQ4: Quality of Short-Horizon Estimates on Performance-Based Student Subgroups}

\begin{figure}
    \centering
    \includegraphics[width=\linewidth]{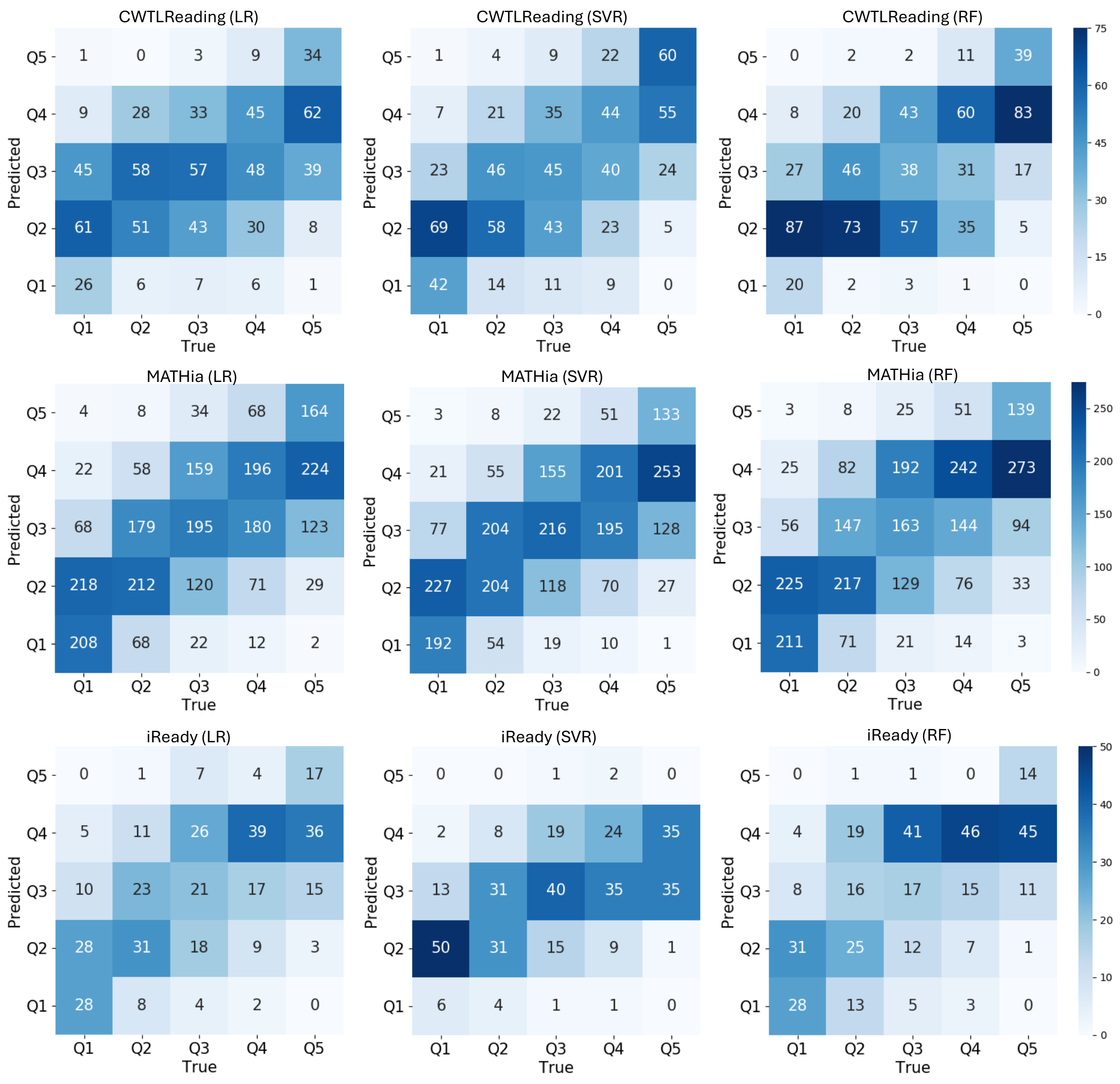}
    \caption{Confusion matrices of the predicted quantile groups using the short horizon log data (Y-axis) and the actual five  subgroups in training set (X-axis), the values are obtained from 5-fold cross validation.}
    \label{fig:heatmap_best}
\end{figure}

As mentioned, though population level prediction accuracy is the most common reported measure,  another important motivation of predicting outcomes using short-horizon data is to help understand and inform additional support and challenge for students in need. We divide students into performance subgroups, by two strategies: 1) using quintile division sorted by their test outcomes (represented by Q1 to Q5 in ascending order); 2) using context-specific binary clusters such as pass and fail (that we refer as `on track' and `not on track', respectively), which are defined by the given assessment. 

\textbf{Quintile subgroups sorted by performance.} Figure~\ref{fig:heatmap_best} shows heat maps of confusion matrices that represent the prediction performance of the three machine learning methods (linear regression, support vector regression and random forest) on three datasets using short horizon. The x-axis shows the true performance subgroups in training set (sorted by post-test scores in ascending order), and the y-axis shows the predicted performance groups. Overall, linear regression (LR) performs better in general at having a stronger diagonal on the confusion matrices. Across all three methods, the accuracy is worst for students predicted in the middle (\textit{i.e.}, Q3 are frequently over- and under-estimated). 

The highest precision is for students at either extreme. For example, for  \texttt{CWTLReading}, when predicting students in the lowest performance quintile (Q1), the best model (random forest) is accurate 77\% of the time  when it predicts someone is going to be Q1, and is accurate 72\% of the time if it predicts someone is going to be Q5. Random forest is also the best at Q1 and Q5 for \texttt{iReady}, with an accuracy of 57\% at predicting Q1, and 88\% at predicting Q5. For \texttt{MATHia} support vector machines has the slight edge over the other models at predictive accuracy for its extreme predictions, reaching an accuracy for predicted Q1 of 70\%, and for Q5 is around 61\%. We note in all cases the other models do worse, but similarly. The models are most inaccurate at predicting quintiles 2 to 5. Note though that the ranges of these quintiles are not evenly spaced (\texttt{CWTLReading}: [0.11, 0.42), [0.42, 0.53), [0.53, 0.62), [0.62, 0.74), [0.74, 1]; \texttt{MATHia}: [0, 0.37), [0.37, 0.48), [0.48, 0.55), [0.55, 0.63), [0.63, 1]; \texttt{iReady}: [0, 0.28), [0.28, 0.42), [0.42, 0.52), [0.52, 0.64), [0.64, 1]). 


While we generally see correlation with the true quantile, the predicted models often overestimate poor performance and underestimate high performance, suggesting it is not yet well-prepared for individual or subgroup level performance predictions. However, the models'   relatively high precision at the extremes suggest they might be able to highlight when students might benefit from additional challenge or support, though further investigation and validation is key before employing such predictions for any important interventions.

\textbf{Predicting subgroups on or off track.} In \texttt{MATHia}, the end-of-year state test also has 5 achievement categories where level 3 to 5 correspond to a passing score. Thus, we consider cluster students into two subgroups, `on track' subgroup whose achievement categories are in level 3 to 5, and `not on track' subgroup whose achievement categories are in level 1 to 2. The prediction process is conducted over the original achievement categories and then we calculate a confusion matrix based on the two clusters. Using the short horizon (2hr) log data, we find that the model correctly classifies as "on track" 1541 students, correctly classifies as "not on track" 423 students, and misclassifies 680 students. In particular, the model has low 
recall for the `not on track' subgroup (0.44, true predicted `not on track' divided by actual `not on track'), indicating it misses a large number of students who are not on track. On the plus side, similar to our results for the quintile analysis, the model has reasonable precision: if it predicts a student is not on track, 74\% of those instances did have a post assessment that indicates they are not on track.



\subsection{RQ5: The Effects of Pre-Assessment Scores}

\begin{table*}[hbtp]
  \caption{Prediction performance using all features extracted from short horizon (5 hours) log data, compared to using pre-test scores and using full log data (Full) for \texttt{CWTLReading}}
  \label{tab:pred_pt_cwtlread}
  \resizebox{\textwidth}{!}{{\begin{tabular}{lllllllllll}
  \toprule
     & \multicolumn{5}{l}{RMSE}           & \multicolumn{5}{l}{$R^2$}                 \\
    & Short & Pre-Test & Short+Pre-Test & Full & Full+Pre-Test & Short & Pre-Test & Short+Pre-Test & Full & Full+Pre-Test \\
    \midrule
Baseline & \multicolumn{5}{l}{0.38 (0.013)} & \multicolumn{5}{l}{0.00 (0.)} \\
LR & 0.14 (0.018) & 0.14 (0.02) & 0.13 (0.009) & 0.16 (0.005) & 0.14 (0.009) & 0.34 (0.194) & 0.39 (0.097) & 0.44 (0.06) & 0.18 (0.05) & 0.28 (0.058) \\
SVR & 0.14 (0.002) & 0.13 (0.002) & 0.13 (0.006) & 0.19 (0.007) & 0.17 (0.006) & 0.38 (0.006) & 0.4 (0.053) & 0.44 (0.048) & 0.1 (0.037) & 0.2 (0.034) \\
RF & 0.13 (0.003) & 0.14 (0.008) & 0.13 (0.005) & 0.16 (0.006) & 0.15 (0.007) & 0.4 (0.017) & 0.39 (0.044) & 0.42 (0.032) & 0.2 (0.064) & 0.27 (0.054) \\
    \bottomrule
  \end{tabular}}
  }
\end{table*}

\begin{table*}[hbtp]
  \caption{Prediction performance using all features extracted from short horizon (3 hours) log data, compared to using pre-test scores and using full log data (Full) for \texttt{iReady}}
  \label{tab:pred_pt_iready}
  \resizebox{\textwidth}{!}{{\begin{tabular}{lllllllllll}
  \toprule
     & \multicolumn{5}{l}{RMSE}           & \multicolumn{5}{l}{$R^2$}                 \\
    & Short & Pre-Test & Short+Pre-Test & Full & Full+Pre-Test & Short & Pre-Test & Short+Pre-Test & Full & Full+Pre-Test \\
    \midrule
Baseline & \multicolumn{5}{l}{0.21 (0.003)} & \multicolumn{5}{l}{0.00 (0.)} \\
LR & 0.17 (0.012) & 0.12 (0.01) & 0.12 (0.009) & 0.17 (0.017) & 0.12 (0.01) & 0.37 (0.076) & 0.65 (0.044) & 0.65 (0.056) & 0.35 (0.085) & 0.64 (0.059) \\
SVR & 0.17 (0.022) & 0.12 (0.01) & 0.12 (0.009) & 0.15 (0.012) & 0.12 (0.009) & 0.37 (0.134) & 0.66 (0.053) & 0.66 (0.057) & 0.47 (0.078) & 0.68 (0.049) \\
RF & 0.16 (0.009) & 0.15 (0.005) & 0.13 (0.008) & 0.15 (0.011) & 0.12 (0.009) & 0.41 (0.061) & 0.54 (0.055) & 0.64 (0.059) & 0.47 (0.073) & 0.65 (0.064) \\
    \bottomrule
  \end{tabular}}
  }
\end{table*}

\textbf{Prediction over population using pre-assessment scores combining with a set of log data features.} Tables~\ref{tab:pred_pt_cwtlread} \& ~\ref{tab:pred_pt_iready} show: 1) using pre-test scores only; 2) combining pre-test with log data features on short horizon; and 3) combining pre-test with full log data features. In general, pre-assessment or pre-test score is a powerful indicator for long-term outcomes, that can be stand-alone and outperforms using log data features only. With pre-test scores, all models show similar performance with RMSE around 0.14 for \texttt{CWTLReading} and around 0.12 for \texttt{iReady}. And RMSE is not significantly improved compared to using log data features in \texttt{CWTLReading}. But $R^2$ values are higher when pre-test score is included, peaking at 0.44 for LR and SVR in the Short+Pre-Test (\texttt{CWTLReading}) and 0.66 for SVR in Short+Pre-Test (\texttt{iReady}), which indicates a better fit model when pre-test scores are integrated. Overall, unsurprizingly, it is beneficial to include a pre-test score in the model if available.

We further investigate the effects of combining single feature with pre-test scores. Table~\ref{tab:single-feature-pt} shows the prediction results on \texttt{CWTLReading} and \texttt{iReady} using LR. Pre-test can substantially enhance LR's performance with single feature. And even when two features have differed effects on model performance (\textit{e.g.}, 0.197 RMSE using PS only vs. 0.179 RMSE using AA only), combining with pre-test could fill the gap by improving prediction performance to the similar level (\textit{e.g.}, 0.123 RMSE using PS+Pre-Test vs. 0.122 RMSE using AA+Pre-Test). Those indicate the effectiveness of pre-test scores in prediction.

\begin{table*}[hbtp]
  \caption{Performance of LR trained using single feature (\textit{i.e.}, \textit{perc\_success\_problem} (PS) and \textit{avg\_attempts\_per\_problem} (AA)) combined with pre-test scores on the two datasets using short horizon.}
\label{tab:single-feature-pt}
 \resizebox{\textwidth}{!}{{\begin{tabular}{lllllllll}
  \toprule
  & \multicolumn{4}{l}{\texttt{CWTLReading}} & \multicolumn{4}{l}{\texttt{iReady}} \\
     & PS & AA & PS+Pre-Test & AA+Pre-Test & PS & AA & PS+Pre-Test & AA+Pre-Test \\
     \midrule
     RMSE & 0.177 (0.007) & 0.177 (0.007) & 0.132 (0.001) & 0.131 (0.002) & 0.197 (0.012) & 0.179 (0.01) & 0.123 (0.009) & 0.122 (0.009) \\ 
     $R^2$ & 0.1 (0.05) & 0.03 (0.01) & 0.4 (0.024) & 0.41 (0.025) & 0.074 (0.053) & 0.236 (0.079) & 0.646 (0.044) & 0.651 (0.043) \\
    \bottomrule
  \end{tabular}}
  }
\end{table*}

\begin{figure}
    \centering
    \includegraphics[width=.78\linewidth]{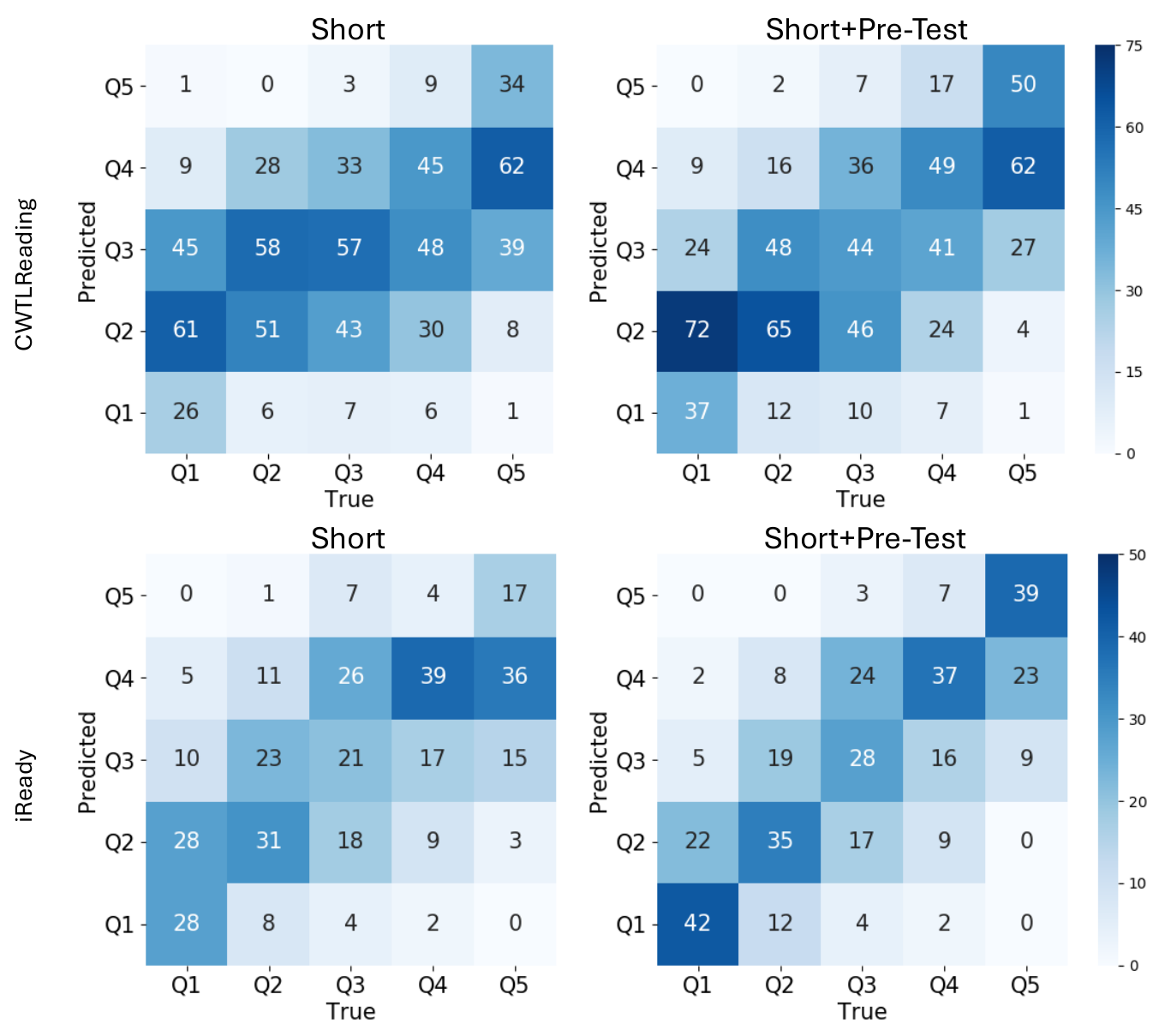}
    \caption{Confusion matrices of the predicted quantile groups using the short horizon log data (Y-axis) and the actual fifth subgroups in training set (X-axis) on \texttt{CWTLReading} and \texttt{iReady} with (Short+Pre-Test) and without (Short) pre-test scores, the values are obtained from LR with 5-fold cross validation.}
    \label{fig:heatmap_pt}
\end{figure}

\textbf{Prediction over subgroups using pre-assessment scores combining with a set of log data features.}
We calculate confusion matrices on quintile division of subgroups using both with and without pre-test scores on selected short horizon. Figure~\ref{fig:heatmap_pt} shows the heat maps of the confusion matrices for \texttt{CWTLReading} and \texttt{iReady}. For lower performers Q1 and Q2 in \texttt{CWTLReading}, using log data features with pretest (Short+Pre-Test) shows more concentrated on overestimating cases (with less on predicting Q1 as Q3-Q5). And for highest performers Q5, Short+Pre-Test shows more concentrated on underestimating cases (with less on predicting Q5 as Q1-Q3). For lowest performers Q1 in \texttt{iReady}, similar to what we observe from \texttt{CWTLReading}, introducing pre-test to log data features helps LR less overestimate. And for highest performers Q5 in \texttt{iReady}, Pre-Test also help LR less underestimate the performers as Q1-Q3.

To summarize, if pre-test score is available, one can use that alone or combine it with a single feature. If we don't have pre-test score and only have log data, there is still a gain to using more than one feature extracted from log data.

\section{Discussions}

From the performance subgroup analysis, we observed that the three machine learning models are not highly accurate at classifying students' likely quintile assessment performance or binary "on track" measure, but the precision of the models is quite good when they suggest someone is in danger of performing very poorly, or is likely to perform very well. This suggests interesting future directions for creating systems that could alert educators or provide additional support or challenge in this setting. Another interesting direction is to see if there are additional features or models that could further improve the predictive accuracy. However, before investigating such, we strongly believe that additional testing and validation would be crucial, particularly to avoid accidentally causing worse outcomes for students, should they be labeled inaccurately or if some teachers or other stakeholders deprioritize such students because they think those students are expected to fail. We also note that stakeholders may wish to understand and interpret the predictions of such models. While linear regression is the most interpretable approach in this work, the others models require additional effort to become interpretable to stakeholders.

In this study, we use the fixed usage time to define short horizons, \textit{i.e.}, we use accumulative interaction time of each student until they reach a threshold of short horizon usage  (\textit{e.g.}, 1 hour). We do not consider a potential usage time situation if some students share the same potential usage time but they do not have productive usage during that time, which may be another important indicator to consider for long term outcomes. In addition, we note that  fixed usage time may result in varied course content coverage across students. For example, imagine if the length of a class is 3 hours (\textit{i.e.}, the potential usage time is 3 hours), but some students only actively engage for 2 hours. If we set a fixed usage time as 3 hours, the considered data of those students will contain their interactions within the 1st hour of the next class. These issues would be an interesting direction for further work. 

We also note that when additional features about student demographics are available, or specific more detailed information about the educational product (such as knowledge components), this information can be used to potentially further enhance predictive accuracy~\citep{ritter2013predicting,zheng2019using}. We did not consider those features in this study, because they were not available and may not always be available in the future, and because our primary interest was a general study using exploring how short-term features that can be extracted across many educational contexts can be used to predict on long-term external assessments.

\section{Conclusions, Limitations, \& Future Work}

In this study, we investigated the efficacy of using short-horizon log data to predict long-term educational outcomes across multiple learning platforms and contexts. Our results show that machine learning models like linear regression and random forest can use data from 2 to 5 hours of educational technology usage (as recorded in three datasets) to provide a useful predictor of long term external outcomes taken months later, with similar performance to using log data from the entire (multi-month) usage period. We find that percentage of success problems and average attempts across problems are generally important predictive features, though more complex machine learning models offer additional predictive power. We also find that our prediction models are not sufficient to provide highly accurate individual predictions, but that they have sufficient precision on predicting those most likely to severely struggle on the post-test, that further investigation could be warranted to see if such information is reliable enough to be used in such tutoring systems, and/or provided as potential information to teachers. Moreover, we find that such short-term data often offers similar predictive power to using pre-test/assessments, and occasionally the combination of both offers additional benefit. As pre-test may not always be available, or may take time away from instruction and practice, short-horizon log data may be a useful tool to consider in predicting long-term external outcomes.


\acks{This work was supported in part by a Stanford Human centered Artificial Intelligence (HAI) seed grant. The authors would also like to express their thanks to the children, school personnel and WCA teams involved in Can’t Wait to Learn programme and research implementation for their data and time.}

\vskip 0.2in
\bibliography{main}

\begin{thebibliography}{37}
\providecommand{\natexlab}[1]{#1}
\providecommand{\url}[1]{\texttt{#1}}
\expandafter\ifx\csname urlstyle\endcsname\relax
  \providecommand{\doi}[1]{doi: #1}\else
  \providecommand{\doi}{doi: \begingroup \urlstyle{rm}\Url}\fi

\bibitem[Acharya and Sinha(2014)]{acharya2014early}
Anal Acharya and Devadatta Sinha.
\newblock Early prediction of students performance using machine learning techniques.
\newblock \emph{International Journal of Computer Applications}, 107\penalty0 (1):\penalty0 37--43, 2014.

\bibitem[Ahadi et~al.(2015)Ahadi, Lister, Haapala, and Vihavainen]{ahadi2015exploring}
Alireza Ahadi, Raymond Lister, Heikki Haapala, and Arto Vihavainen.
\newblock Exploring machine learning methods to automatically identify students in need of assistance.
\newblock In \emph{Proceedings of the eleventh annual International Conference on International Computing Education Research}, pages 121--130, 2015.

\bibitem[Alhazmi and Sheneamer(2023)]{alhazmi2023early}
Essa Alhazmi and Abdullah Sheneamer.
\newblock Early predicting of students performance in higher education.
\newblock \emph{IEEE Access}, 11:\penalty0 27579--27589, 2023.

\bibitem[Anozie and Junker(2006)]{anozie2006predicting}
Nathaniel Anozie and Brian~W Junker.
\newblock Predicting end-of-year accountability assessment scores from monthly student records in an online tutoring system.
\newblock Educational Data Mining: Papers from the AAAI Workshop. Menlo Park, CA: AAAI~…, 2006.

\bibitem[Athey et~al.(2019)Athey, Chetty, Imbens, and Kang]{athey2019surrogate}
Susan Athey, Raj Chetty, Guido~W Imbens, and Hyunseung Kang.
\newblock The surrogate index: Combining short-term proxies to estimate long-term treatment effects more rapidly and precisely.
\newblock Technical report, National Bureau of Economic Research, 2019.

\bibitem[Ayers and Junker(2008)]{ayers2008irt}
Elizabeth Ayers and Brian Junker.
\newblock Irt modeling of tutor performance to predict end-of-year exam scores.
\newblock \emph{Educational and Psychological Measurement}, 68\penalty0 (6):\penalty0 972--987, 2008.

\bibitem[Beck and Gong(2013)]{beck2013wheel}
Joseph~E Beck and Yue Gong.
\newblock Wheel-spinning: Students who fail to master a skill.
\newblock In \emph{Artificial Intelligence in Education: 16th International Conference, AIED 2013, Memphis, TN, USA, July 9-13, 2013. Proceedings 16}, pages 431--440. Springer, 2013.

\bibitem[Brown et~al.(2023)Brown, Farag, Hussein Abd~Alla, Radford, Miller, Neijenhuijs, Stubb{\'e}, de~Hoop, Abdullatif~Abbadi, Turner, et~al.]{brown2023can}
Felicity~L Brown, Alawia~I Farag, Faiza Hussein Abd~Alla, Kate Radford, Laura Miller, Koen Neijenhuijs, Hester Stubb{\'e}, Thomas de~Hoop, Ahmed Abdullatif~Abbadi, Jasmine~S Turner, et~al.
\newblock Can’t wait to learn: A quasi-experimental mixed-methods evaluation of a digital game-based learning programme for out-of-school children in sudan.
\newblock \emph{Journal of Development Effectiveness}, 15\penalty0 (3):\penalty0 320--341, 2023.

\bibitem[Carter and Hundhausen(2017)]{carter2017using}
Adam~Scott Carter and Christopher~David Hundhausen.
\newblock Using programming process data to detect differences in students' patterns of programming.
\newblock In \emph{Proceedings of the 2017 ACM SIGCSE Technical Symposium on Computer Science Education}, pages 105--110, 2017.

\bibitem[Castro-Wunsch et~al.(2017)Castro-Wunsch, Ahadi, and Petersen]{castro2017evaluating}
Karo Castro-Wunsch, Alireza Ahadi, and Andrew Petersen.
\newblock Evaluating neural networks as a method for identifying students in need of assistance.
\newblock In \emph{Proceedings of the 2017 ACM SIGCSE technical symposium on computer science education}, pages 111--116, 2017.

\bibitem[Emerson et~al.(2019)Emerson, Rodr{\'\i}guez, Mott, Smith, Min, Boyer, Smith, Wiebe, and Lester]{emerson2019predicting}
Andrew Emerson, Fernando~J Rodr{\'\i}guez, Bradford Mott, Andy Smith, Wookhee Min, Kristy~Elizabeth Boyer, Cody Smith, Eric Wiebe, and James Lester.
\newblock Predicting early and often: Predictive student modeling for block-based programming environments.
\newblock In \emph{Proceedings of The 12th International Conference on Educational Data Mining (EDM 2019)}, volume~39, page~48. ERIC, 2019.

\bibitem[Feng et~al.(2006)Feng, Heffernan, and Koedinger]{feng2006predicting}
Mingyu Feng, Neil~T Heffernan, and Kenneth~R Koedinger.
\newblock Predicting state test scores better with intelligent tutoring systems: developing metrics to measure assistance required.
\newblock In \emph{International conference on intelligent tutoring systems}, pages 31--40. Springer, 2006.

\bibitem[Gao et~al.(2021)Gao, Marwan, and Price]{gao2021early}
Ge~Gao, Samiha Marwan, and Thomas~W Price.
\newblock Early performance prediction using interpretable patterns in programming process data.
\newblock In \emph{Proceedings of the 52nd ACM technical symposium on computer science education}, pages 342--348, 2021.

\bibitem[Gao et~al.(2022)Gao, Gao, Yang, Pajic, and Chi]{gao2022reinforcement}
Ge~Gao, Qitong Gao, Xi~Yang, Miroslav Pajic, and Min Chi.
\newblock A reinforcement learning-informed pattern mining framework for multivariate time series classification.
\newblock In \emph{31st International Joint Conference on Artificial Intelligence (IJCAI)}, 2022.

\bibitem[Grover et~al.(2017)Grover, Basu, Bienkowski, Eagle, Diana, and Stamper]{grover2017framework}
Shuchi Grover, Satabdi Basu, Marie Bienkowski, Michael Eagle, Nicholas Diana, and John Stamper.
\newblock A framework for using hypothesis-driven approaches to support data-driven learning analytics in measuring computational thinking in block-based programming environments.
\newblock \emph{ACM Transactions on Computing Education (TOCE)}, 17\penalty0 (3):\penalty0 1--25, 2017.

\bibitem[Helal et~al.(2019)Helal, Li, Liu, Ebrahimie, Dawson, and Murray]{helal2019identifying}
Sumyea Helal, Jiuyong Li, Lin Liu, Esmaeil Ebrahimie, Shane Dawson, and Duncan~J Murray.
\newblock Identifying key factors of student academic performance by subgroup discovery.
\newblock \emph{International Journal of Data Science and Analytics}, 7:\penalty0 227--245, 2019.

\bibitem[Hohnhold et~al.(2015)Hohnhold, O'Brien, and Tang]{hohnhold2015focusing}
Henning Hohnhold, Deirdre O'Brien, and Diane Tang.
\newblock Focusing on the long-term: It's good for users and business.
\newblock In \emph{Proceedings of the 21th ACM SIGKDD International Conference on Knowledge Discovery and Data Mining}, pages 1849--1858, 2015.

\bibitem[Jadud(2006)]{jadud2006methods}
Matthew~C Jadud.
\newblock Methods and tools for exploring novice compilation behaviour.
\newblock In \emph{Proceedings of the second international workshop on Computing education research}, pages 73--84, 2006.

\bibitem[Joshi et~al.(2014)Joshi, Fancsali, Ritter, Nixon, and Berman]{joshi2014generalizing}
Ambarish Joshi, Stephen Fancsali, Steven Ritter, Tristan Nixon, and Susan Berman.
\newblock Generalizing and extending a predictive model for standardized test scores based on cognitive tutor interactions.
\newblock In \emph{Educational Data Mining 2014}. Citeseer, 2014.

\bibitem[Leon et~al.(2024)Leon, Nie, Chandak, and Brunskill]{leon2024estimating}
Amelia Leon, Allen Nie, Yash Chandak, and Emma Brunskill.
\newblock Estimating the causal treatment effect of unproductive persistence.
\newblock In \emph{Proceedings of the 14th Learning Analytics and Knowledge Conference}, pages 843--849, 2024.

\bibitem[Liao et~al.(2019)Liao, Zingaro, Thai, Alvarado, Griswold, and Porter]{liao2019robust}
Soohyun~Nam Liao, Daniel Zingaro, Kevin Thai, Christine Alvarado, William~G Griswold, and Leo Porter.
\newblock A robust machine learning technique to predict low-performing students.
\newblock \emph{ACM Transactions on Computing Education (TOCE)}, 19\penalty0 (3):\penalty0 1--19, 2019.

\bibitem[Mao(2019)]{mao2019one}
Ye~Mao.
\newblock One minute is enough: Early prediction of student success and event-level difficulty during novice programming tasks.
\newblock In \emph{In: Proceedings of the 12th International Conference on Educational Data Mining (EDM 2019)}, 2019.

\bibitem[Mao et~al.(2020)Mao, Marwan, Price, Barnes, and Chi]{mao2020time}
Ye~Mao, Samiha Marwan, Thomas~W Price, Tiffany Barnes, and Min Chi.
\newblock What time is it? student modeling needs to know.
\newblock In \emph{In proceedings of the 13th International Conference on Educational Data Mining}, 2020.

\bibitem[Pardos et~al.(2006)Pardos, Heffernan, Anderson, Heffernan, and Schools]{pardos2006using}
Zachary~A Pardos, Neil~T Heffernan, Brigham Anderson, Cristina~L Heffernan, and Worcester~Public Schools.
\newblock Using fine-grained skill models to fit student performance with bayesian networks.
\newblock In \emph{Workshop in Educational Data Mining held at the 8th International Conference on Intelligent Tutoring Systems. Taiwan}, 2006.

\bibitem[Pardos et~al.(2014)Pardos, Baker, San~Pedro, Gowda, and Gowda]{pardos2014affective}
Zachary~A Pardos, Ryan~SJD Baker, Maria~OCZ San~Pedro, Sujith~M Gowda, and Supreeth~M Gowda.
\newblock Affective states and state tests: Investigating how affect and engagement during the school year predict end-of-year learning outcomes.
\newblock \emph{Journal of Learning Analytics}, 1\penalty0 (1):\penalty0 107--128, 2014.

\bibitem[Piech et~al.(2012)Piech, Sahami, Koller, Cooper, and Blikstein]{piech2012modeling}
Chris Piech, Mehran Sahami, Daphne Koller, Steve Cooper, and Paulo Blikstein.
\newblock Modeling how students learn to program.
\newblock In \emph{Proceedings of the 43rd ACM technical symposium on Computer Science Education}, pages 153--160, 2012.

\bibitem[Porter et~al.(2014)Porter, Zingaro, and Lister]{porter2014predicting}
Leo Porter, Daniel Zingaro, and Raymond Lister.
\newblock Predicting student success using fine grain clicker data.
\newblock In \emph{Proceedings of the tenth annual conference on International computing education research}, pages 51--58, 2014.

\bibitem[Prihar(2023)]{prihar2023effective}
Ethan Prihar.
\newblock Effective evaluation of online learning interventions with surrogate measures.
\newblock In \emph{In The Proceedings of the 16th International Conference on Educational Data Mining}, 2023.

\bibitem[Ritter et~al.(2013)Ritter, Joshi, Fancsali, and Nixon]{ritter2013predicting}
Steve Ritter, Ambarish Joshi, Stephen Fancsali, and Tristan Nixon.
\newblock Predicting standardized test scores from cognitive tutor interactions.
\newblock In \emph{Educational Data Mining 2013}, 2013.

\bibitem[Ritter et~al.(2007)Ritter, Anderson, Koedinger, and Corbett]{ritter2007cognitive}
Steven Ritter, John~R Anderson, Kenneth~R Koedinger, and Albert Corbett.
\newblock Cognitive tutor: Applied research in mathematics education.
\newblock \emph{Psychonomic bulletin \& review}, 14:\penalty0 249--255, 2007.

\bibitem[Tran et~al.(2023)Tran, Yeatman, Burkhardt, Ma, Mitchell, Yablonski, Gijbels, Townley-Flores, and Richie-Halford]{tran2023development}
Jasmine~Elizabeth Tran, Jason Yeatman, Amy Burkhardt, Wanjing~Anya Ma, Jamie Mitchell, Maya Yablonski, Liesbeth Gijbels, Carrie Townley-Flores, and Adam Richie-Halford.
\newblock Development and validation of a rapid online sentence reading efficiency assessment.
\newblock 2023.

\bibitem[Walonoski and Heffernan(2006)]{walonoski2006detection}
Jason~A Walonoski and Neil~T Heffernan.
\newblock Detection and analysis of off-task gaming behavior in intelligent tutoring systems.
\newblock In \emph{Intelligent Tutoring Systems: 8th International Conference, ITS 2006, Jhongli, Taiwan, June 26-30, 2006. Proceedings 8}, pages 382--391. Springer, 2006.

\bibitem[Wang et~al.(2022)Wang, Sharma, Xu, Badam, Sun, Richardson, Chung, Chi, and Chen]{wang2022surrogate}
Yuyan Wang, Mohit Sharma, Can Xu, Sriraj Badam, Qian Sun, Lee Richardson, Lisa Chung, Ed~H Chi, and Minmin Chen.
\newblock Surrogate for long-term user experience in recommender systems.
\newblock In \emph{Proceedings of the 28th ACM SIGKDD conference on knowledge discovery and data mining}, pages 4100--4109, 2022.

\bibitem[Watson et~al.(2013)Watson, Li, and Godwin]{watson2013predicting}
Christopher Watson, Frederick~WB Li, and Jamie~L Godwin.
\newblock Predicting performance in an introductory programming course by logging and analyzing student programming behavior.
\newblock In \emph{2013 IEEE 13th international conference on advanced learning technologies}, pages 319--323. IEEE, 2013.

\bibitem[Zhang et~al.(2019)Zhang, Huang, Wang, Lu, Fang, Stamper, Fancsali, Holstein, and Aleven]{zhang2019early}
Chuankai Zhang, Yanzun Huang, Jingyu Wang, Dongyang Lu, Weiqi Fang, John Stamper, Stephen Fancsali, Kenneth Holstein, and Vincent Aleven.
\newblock Early detection of wheel spinning: Comparison across tutors, models, features, and operationalizations.
\newblock \emph{International Educational Data Mining Society}, 2019.

\bibitem[Zhang et~al.(2023)Zhang, Zhao, Le, and Kallus]{zhang2023evaluating}
Vickie Zhang, Michael Zhao, Anh Le, and Nathan Kallus.
\newblock Evaluating the surrogate index as a decision-making tool using 200 a/b tests at netflix.
\newblock \emph{arXiv e-prints}, pages arXiv--2311, 2023.

\bibitem[Zheng et~al.(2019)Zheng, Fancsali, Ritter, and Berman]{zheng2019using}
Guoguo Zheng, Stephen~Edward Fancsali, Steven Ritter, and Susan Berman.
\newblock Using instruction-embedded formative assessment to predict state summative test scores and achievement levels in mathematics.
\newblock \emph{Journal of Learning Analytics}, 6\penalty0 (2):\penalty0 153--174, 2019.

\end{thebibliography}

\end{document}